\documentclass[letterpaper, 10 pt, conference]{ieeeconf}  

\IEEEoverridecommandlockouts                              

\usepackage{graphics} 
\usepackage{epsfig} 
\usepackage{amsmath} 
\usepackage{amssymb}  
\usepackage{xcolor}
\usepackage{nicefrac}
\usepackage{siunitx}
\def\BibTeX{{\rm B\kern-.05em{\sc i\kern-.025em b}\kern-.08em
    T\kern-.1667em\lower.7ex\hbox{E}\kern-.125emX}}

\usepackage{tabularray}
\UseTblrLibrary{amsmath} 
\usepackage{caption}
\usepackage{subcaption}

\usepackage[noconfig,amsmath,thmmarks]{ntheorem}
\theoremnumbering{arabic}
\theoremstyle{plain}
\RequirePackage{latexsym}
\theorembodyfont{\itshape}
\theoremheaderfont{\normalfont\bfseries}
\theoremseparator{}

\usepackage{tikz}
\usepackage{pgfplots}
\usepackage{cite}
\usepackage{makecell}

\newcommand{\bm}[1] {\boldsymbol{#1} }

\title{\LARGE \bf
Study on Human-Variability-Respecting Optimal Control Affecting Human Interaction Experience
}

\author{Sean Kille$^{1}$, Balint Varga$^{1}$, Sören Hohmann$^{1}$
\thanks{$^{1}$All authors are with the Institute of Control Systems (IRS), Karlsruhe Institute of Technology (KIT), 76131 Karlsruhe, Germany. Corresponding author is {\tt\small sean.kille@kit.edu}.}%
}

\begin{document}

\maketitle
\thispagestyle{empty}
\pagestyle{empty}

\begin{abstract}
Broad application of human-machine interaction (HMI) demands advanced and human-centered control designs for the machine's automation. Human natural motor action shows stochastic behavior, which has so far not been respected in HMI control designs. Using a previously presented novel human-variability-respecting optimal controller we present a study design which allows the investigation of respecting human natural variability and its effect on human interaction experience. Our approach is tested in simulation based on an identified real human subject and presents a promising approach to be used for a larger subject study.  

\end{abstract}


\section{Introduction}
With Human-Machine Interaction being applied to an increasing number of use-cases in various domains, further development of the machine's control is of strong interest. One aspect of the machines enhancement is to design it more human-centric. This requires a consideration of human natural behavior, which naturally shows movement variability in repetitive movements~\cite{Abend.1982, Harris.1998}. Looking at literature from neuroscience on motor control, stochastic optimal control models best describe this human motor behavior~\cite{Karg.2022}, the main representative being the linear-quadratic sensorimotor (LQS) model. However, when focusing on the state of the art in physical Human-Robot Interaction, few automation approaches are based on a stochastic human model and none to our knowledge explicitly incorporate human natural variability in their design. This mismatch leads to two questions: Q1) How to design an automation that explicity considers human stochasticity and allows for high variability in task-irrelevant areas? Q2) How does the human experience a freedom or restriction of task-irrelevant variability?
We addressed Q1 in our previous contribution~\cite{Kille.2024}, introducing a Human-Variability-Respecting Optimal Control (HVROC). The contribution of this paper is the presentation of an approach to investigate Q2, using our proposed solution to Q1. We firstly introduce our system model in~\ref{sec:model} and our experimental system in~\ref{sec:experimentalSystem}. Based on measurements of subjects interacting with this system, we use an inverse stochastic optimal control algorithm to identify human cost function and noise process parameters (\ref{sec:identification}). Using these parameters, we parametrize individual control laws as proposed in our previous publication (\ref{sec:automation}). The study procedure is then explained using an identified subject and a control law in simulation (\ref{sec:procedure}).


\section{Materials and Methods} 
This section introduces our human-machine system model and the experimental system setup, before explaining our identification procedure and proposed automation design. 

\subsection{Human-Machine System Model}\label{sec:model}

We introduce a system which is acted upon by both a human and an automation. The automation is expected to act deterministically, while we model the human motor behavior as being affected by additive and multiplicative noise processes in action and perception as presented by~\cite{Todorov.2005}:

\begin{align}
    \bm{x}_{t+1}=& \bm{A} \bm{x}_t  + \bm{B}_{\mathrm{A}} \bm{u}_{\mathrm{A},t}+ \bm{B}_{\mathrm{H}} \bm{u}_{\mathrm{H},t} \notag \\
    &+ \bm{\Sigma}^{\bm{\alpha}} \bm{\alpha}_t + \sum_i^c \varepsilon_t^{(i)} \bm{C}_i \bm{u}_{\mathrm{H},t},  \label{eq:system} \\
    \bm{y}_{\mathrm{H},t} &= \bm{H}_{\mathrm{H}} \bm{x}_t + 
    \bm{\Sigma}^{\bm{\beta}} \bm{\beta}_{t} + \sum_{i}^{d} \epsilon_t^{(i)} \bm{D}_i \bm{x}_t, \label{eq:sysOutH} \\
    \bm{y}_{\mathrm{A},t} &= \bm{H}_{\mathrm{A}} \bm{x}_t. \label{eq:sysOutA}
\end{align}
With $\bm{x} \in \mathbb{R}^{n}$ we denote the system state, with $\bm{y}_{\mathrm{H}} \in \mathbb{R}^{r_\mathrm{H}}$ and $\bm{y}_{\mathrm{A}} \in \mathbb{R}^{r_\mathrm{A}}$ we denote the human and automation perception, respectively. $\bm{u}_{\mathrm{H}} \in \mathbb{R}^{m_H}$ and $\bm{u}_{\mathrm{A}}\in\mathbb{R}^{m_A}$ describe the control variables of the human and automation.
The human action and perception are overlaid with an additive standard white Gaussion noise process $\bm{\alpha}$ and $\bm{\beta}$ with the scaling parameters $\bm{\Sigma}^{\bm{\alpha}}$ and $\bm{\Sigma}^{\bm{\beta}}$. Additionally  we include the control-dependent noise process  $\sum_i \varepsilon_t^{(i)} \bm{C}_i \bm{u}_{\mathrm{H},t}$ and state-dependend noise process$ \sum_i \epsilon_t^{(i)} \bm{D}_i \bm{x}_t$ with $\bm{\varepsilon} = \begin{bmatrix} \varepsilon_t^{(1)}  \dots \varepsilon_t^{(c)} \end{bmatrix}^\intercal$ and  $\bm{\epsilon} = \begin{bmatrix} \epsilon_t^{(1)}  \dots \epsilon_t^{(d)} \end{bmatrix}^\intercal$ denoting standard white Gaussian noise. 

\subsubsection{Human Action}
We model the human movement using a linear-quadratic sensorimotor model as introduced by~\cite{Todorov.2005}. It assumes that the human action aims at minimizing the cost function
\begin{align}\label{eq:costHuman}
    J_\mathrm{H} = &  \mathrm{E} \biggl\{
    \bm{x}_N^\intercal \bm{Q}_{\mathrm{H},N}\bm{x}_{N} \notag \\ 
    & + \sum_{t=0}^{N-1} \left( \bm{x}_t^\intercal \bm{Q}_{\mathrm{H},t} \bm{x}_t + \bm{u}_{\mathrm{H},t}^\intercal \bm{R}_{\mathrm{H}} \bm{u}_{\mathrm{H},t} \right) \biggr\},
\end{align}
using a linear time-variant control law of the form: $\bm{u}_\mathrm{H} = \bm{L}_{\mathrm{H}} \bm{x}$.

\subsection{Experimental System}\label{sec:experimentalSystem}

As our experimental system we use a 1D simulated  point mass-damper system, using the following dynamic equations:

\begin{subequations}
    \begin{align}
        p_{x,t+1}&= p_{x,t} + \Delta t \; \Dot{p}_{x,t}, \label{eq:sys_p}\\
        \Dot{p}_{x,t+1}&= \left( 1- \Delta t \frac{d}{m} \right) \Dot{p}_{x,t} + \frac{\Delta t}{m} f_{x,t}, \label{eq:sys_v} 
    \end{align}
\end{subequations}
with $p_x$ and $\dot{p}_x$ denoting the position and velocity of the point mass $m$ with the damping factor $d$. 
Furthermore we model the human muscle force $f_{x}$ to be the output of a second-order linear filter $g_x$ with the human neural activation $u_{\mathrm{H},x}$ as input, adhering to the model of~\cite{Todorov.2005}: 
\begin{subequations}
    \begin{align}
        f_{x,t+1} &= \left( 1 - \frac{\Delta t}{\tau_2} \right) f_{x,t} + \frac{\Delta t}{\tau_2} g_{x,t} + u_{\mathrm{A},x,t},  \label{eq:sys_f} \\ 
         g_{x,t+1} &= \left( 1 - \frac{\Delta t}{\tau_1} \right) g_{x,t} + \frac{\Delta t}{\tau_1} u_{\mathrm{H},x,t}. \label{eq:sys_g}
    \end{align}
\end{subequations}

An automation's input $u_{\mathrm{A},x}$ adds to the overall exerted force $f_x$. The system state is defined as $\bm{x} = \begin{bmatrix} p_x & \dot{p}_x & f_x & g_x & p_{x,\mathrm{ref}} \end{bmatrix}^{\intercal}$.
The manipulated object is simulated as a point-mass with an inertia of $m = \SI{50}{kg}$ and damped with $d = \SI{75}{kg/s}$. 
We constraint the system to be movable in the $x$-dimension only. 
From the dynamic equations, the human system matrices $\bm{A}, \bm{B}_{\mathrm{H}}$ and $\bm{B}_{\mathrm{A}}$ can be derived. 
Regarding the human and automation perception, we set $\bm{H}_\mathrm{H} = 
\begin{bmatrix} \bm{I}_{3 \times 3} & \bm{0}_{3 \times 2}  \end{bmatrix}$ 
and 
$\bm{H}_\mathrm{A} = 
\begin{bmatrix} \bm{I}_{3 \times 2} & \bm{0}_{3 \times 3} \end{bmatrix}$. 

We assume all stochastic processes to be independent. The scaling parameters for the additive noise we defined as: $\bm{\Sigma}^{\bm{\alpha}} = \mathrm{diag} (\begin{bmatrix} \sigma_1 & \sigma_2 & \sigma_3 & \sigma_4 & 0 \end{bmatrix})$ 
and $ \bm{\Sigma}^{\bm{\beta}} = \mathrm{diag} (\begin{bmatrix} \sigma_5 & \sigma_6 & \sigma_7 \end{bmatrix})$. The signal-dependend noise scaling parameters are defined as:
\begin{align}
    \sigma^u \bm{B}_\mathrm{H} \bm{F} = & \sigma_8 \bm{B}_\mathrm{H}, \notag \\
    \sigma^x_1 \bm{H}_\mathrm{H} \bm{G}_1 = & \sigma_9 \bm{H}_\mathrm{H} \mathrm{diag} (\begin{bmatrix} 1 & 0 & 0 & 0 & 0 \end{bmatrix}), \notag \\
    \sigma^x_2 \bm{H}_\mathrm{H} \bm{G}_2 = & \sigma_{9} \bm{H}_\mathrm{H} \mathrm{diag} (\begin{bmatrix} 0 & 1 & 0 & 0 & 0 \end{bmatrix}),\notag \\
    \sigma^x_3 \bm{H}_\mathrm{H} \bm{G}_3 = & \sigma_{9} \bm{H}_\mathrm{H} \mathrm{diag} (\begin{bmatrix} 0 & 0 & 0 & 0 & 1 \end{bmatrix}). \notag
\end{align}

\begin{figure}
    \begin{center}
    \includegraphics[width=6.4cm]{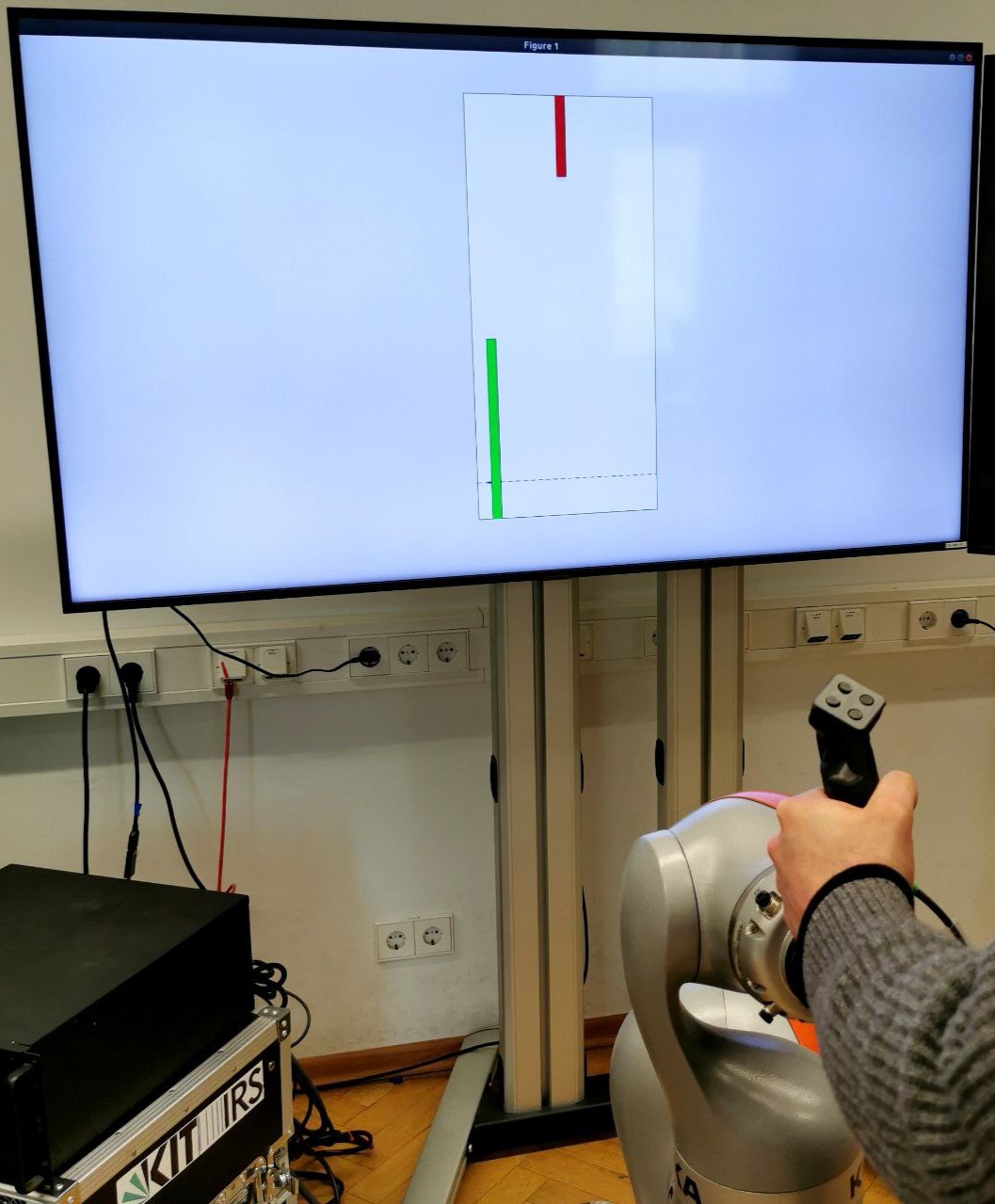}    
    \caption{Experimental setup.} 
    \label{fig:experiment}
    \end{center}
\end{figure}

A user can interact with the simulated experimental system by means of a haptic interface using a robotic arm (KUKA LBR iiwa 14 R820), see~\cite{Braun.2023}. The haptic interface tracks the position of the simulated   point-mass and measures the user's input forces, which serve as a input to the experimental system. The position of the experimental system is additionally visualized using a GUI, which displays the set-positions using falling blocks. The system setup is displayed in Fig.~\ref{fig:experiment}.

\subsection{Identification} \label{sec:identification}
For identification of the model parameters describing human motor behavior, we choose the Inverse Stochastic Optimal Control identification method propsed by~\cite{Karg.2024}. It offers a bi-level approach to determine both the human cost function parameters and noise scaling parameters based on observed human trajectories. The cost-function parameters are combined in the vector $\bm{s}$, with $s_i$ representing the diagonal elements of the $\bm{Q}_{\mathrm{H},N}, \bm{Q}_{\mathrm{H},t}$ and $\bm{R}_{\mathrm{H}}$. The noise parameters are conglomerated in vector $\bm{\sigma}$. For a detailed explanation of the identification algorithm we refer to~\cite{Karg.2024}. 

\subsection{Automation}\label{sec:automation}

As an automation we use our Human-Variability-Respecting Optimal Controller (HVROC) as introduced in~\cite{Kille.2024}. A time-variant optimal control-law and state-estimation is derived, considering human cost function and noise parameters and being optimized to maintain natural human variability considering the effect of the noise parameters on the mean and variability of the resulting trajectories. The controller can be parametrized to result in a joint behavior that either matches the position variability of the human-only interaction or to result in a reduced position variability.

\section{Study Procedure}\label{sec:procedure}
In this section, we present our proposed study procedure and illustrate it in simulation using one identified subject. 

\subsection{Procedure}
Firstly, the human model parameters need to be identified. Therefore, we let the human interact with the experimental system and and record the interaction. Using the algorithm mentioned in~\ref{sec:identification}, the human modeled cost and noise parameters are identified. Based on these, the human-only behavior for the relevant movement(s) is simulated. Using our HVROC code introduced in~\cite{Kille.2024}, an optimal control law respecting the human-only behavior can be determined. We choose two parametrizations, one aiming at mimicking the human-only variability ($\bm{L}_{\mathrm{A, high}}$) and one reducing it ($\bm{L}_{\mathrm{A, low}}$). In two subsequent study runs the participant interacts with the experimental system, being supported by the automations each. We will perform an objective analysis regarding the performance of the joint interaction
and perform a subjective evaluation by assessing the subject's experience using standardized questionnaires.  

\subsection{Numerical example}

To illustrate our approach, we analyze a subject from a preliminary study we conducted and test the developed control law in simulation.

\begin{figure}[h]
    \centering
    \begin{subfigure}[b]{0.475\textwidth}
        \centering
    \resizebox{3.25in}{!}{
    \input{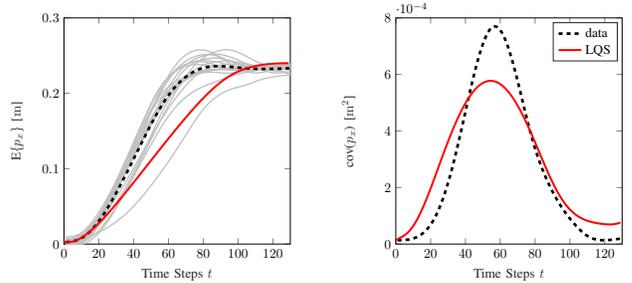}}
    \caption{Ground-truth data and identified LQS model of human subject. Gray depicts the 17 reaching movements, dotted black plots the mean and variance of all repetitions. The LQS model simulation based on identified parameters is depicted in red.  }
    \label{fig:plotGT}
    \end{subfigure}
    \begin{subfigure}[b]{0.475\textwidth}
        \centering
    \resizebox{3.25in}{!}{
    \input{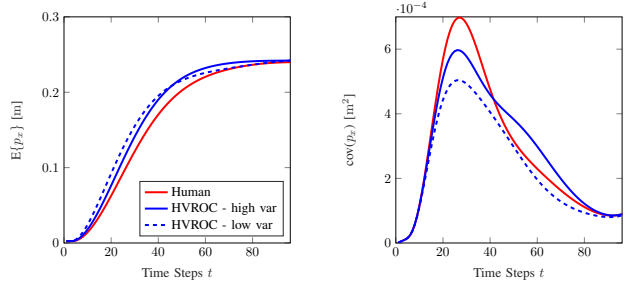}}
    \caption{Simulation result of the HVROC controllers $\bm{L}_{\mathrm{A, high}}$ and $\bm{L}_{\mathrm{A, low}}$ compared to human-only simulated behavior. }
    \label{fig:plotHVROC}
    \end{subfigure}
    \caption{From ground-truth data to simulated interaction of mean and variance of the position $p_x$. Human-only measured data is depicted in gray and dashed black and LQS human simulation behavior is depicted in red. The joint behavior of our proposed \textit{HVROC} with the simulated human is shown in blue; solid and dotted lines represent a high and low desired variability in the task-irrelevant area, respectively.}
    \label{fig:plots}
\end{figure}

Our subject at hand performed $75$ repetitions of a 1-dimensional reaching task, moving a virtual mass between reference points which are located at $T_{x} = \SI{-12}{cm} / \SI{0}{cm} / \SI{12}{cm}$ on the $x$-axis. We choose to analyze the $17$ repetitions from $T_{x} = \SI{-12}{cm}$ to $\SI{12}{cm}$, which are depicted in gray in Fig.~\ref{fig:plotGT}. The mean and variance of the movements are plotted with dashed black lines. Based on the data, the human LQS model parameters are identified using the algorithm presented in~\ref{sec:identification}. We set $\bm{Q}_{\mathrm{H},t} = \bm{0}$ and identify 
\begin{align}
    \bm{R}_{\mathrm{H}} &= 3.43e-9, \notag \\
    \bm{Q}_{\mathrm{H},N} &= \mathrm{diag} (\begin{bmatrix} 1 & 0.253 & 34.2 & 0 & 0 \end{bmatrix}) \notag 
\end{align}
for cost function parameters and 
\begin{align}
    \bm{\sigma}_{1:4} &= \begin{bmatrix} 1.8e-3 & 1.10e-5 & 1.63e-2 & 1.68e-2  \end{bmatrix}, \notag \\
    \bm{\sigma}_{5:7} &= \begin{bmatrix} 1.61e-2 & 4.12e-2 & 1.53e-2 \end{bmatrix}, \notag \\
    \bm{\sigma}_{8:9} &= \begin{bmatrix} 1.25 & 1.4e-3 \end{bmatrix} \notag
\end{align}
for noise parameters. The resulting modeled behavior is depicted in red in Fig.~\ref{fig:plotGT}.

On basis of the identified parameters, we derive two time-variant control laws, $\bm{L}_{\mathrm{A, low}}$ and $\bm{L}_{\mathrm{A, high}}$. The resulting behavior of the simulated human together with the automation is depicted in Fig.~\ref{fig:plotHVROC}. Due to the exemplary nature of this example, we focus on a qualitative evaluation: The simulation shows that $\bm{L}_{\mathrm{A, low}}$ leads to a reduced joint peak variance compared to $\bm{L}_{\mathrm{A, high}}$. Also, the reference point of $p_{x,\mathrm{ref}}=\SI{0.24}{m}$ is reached faster. 

\subsection{Discussion}
When looking at the identified LQS model in Fig.~\ref{fig:plotGT}, an offset between the measured data and the simulated model becomes evident. The simulation behaves slower than the measured mean. Also, the simulated peak variance is about $25\%$ lower than the measured variance. For our proposed study, especially the variance of the human movement is of relevance. Even though an exact match of the measured data is not possible, the magnitudes of both simulation and data are similar and might therefore provide a sufficient basis for controller development.

The simulation results in Fig.~\ref{fig:plotHVROC} show that our approach works as intended but not as distinct as hoped. The two parametrizations lead to a difference in simulated variance, with the low-variance parametrization leading to a variance about $20\%$ than that of the high-variance parametrization. Both reduce the variance compared to the human-only behavior, which stems from the fact that an additional deterministic actor leads to more deterministic behavior. Both controllers lead to a faster reference point reaching and show a similar endpoint variance to the human-only simulation, proving in simulation that a similar or better performance is expected compared to human-only behavior. Overall, the results validate in simulation our solution to Q1 based on a real identified subject.   

The provided results show that a realization of the planned study design should be pursued, in order to validate the preliminary simulation results with human participants. Additionally, with real subject data, the participants experience should be assessed in order to answer Q2. This will provide insights if human stochasticity should be considered in human-centered control designs.   

\section{Conclusion}

In this paper, we adress the observation that no state-of-the-art human-machine interaction approach explicitly incorporates a stochastic model of human variability motor behavior in their control design. We propose a study design that will allow the investigation of how the human experiences a freedom vs. a restriction of his natural variability in an interaction task while being assisted by a robot. The procedure is tested using one real subject data for identification and simulating the resulting interaction. We conclude that the approach is promising and should be validated using a real subject study.






\newpage
\bibliography{1_tex/bibliography.bib}
\bibliographystyle{IEEEtran}


\end{document}